\documentclass[twocolumn,amsmath,amssymb,pra,aps]{revtex4}
\usepackage{color}
\usepackage{multirow} 
\usepackage{graphicx}
\usepackage{epstopdf}
\usepackage{array}
\usepackage[normalem]{ulem}
\usepackage{soul}
\usepackage{booktabs}
\usepackage{natbib}

\usepackage{float}
\usepackage[normalem]{ulem}
\usepackage{amsmath}

\newcolumntype{x}[1]{>{\centering\hspace{0pt}}p{#1}}
\begin{document}
\title{Using electric fields for pulse compression and group velocity control}

\author{Qian Li$^1$}

\email{qian.li@fysik.lth.se}				
\author{Adam Kinos$^1$}	
\author{Axel Thuresson$^2$}	
\author{Lars Rippe$^{1, 3}$}	
\author{Stefan Kr\"oll$^1$}	
\affiliation{$^1$Department of physics, Lund University, P.O. Box 118, SE-22100 Lund, Sweden}
\affiliation{$^2$Theoretical Chemistry, Lund University, P.O. Box 124, SE-22100 Lund, Sweden}
\affiliation{$^3$Flatfrog, Scheelevägen 15, SE-22363 Lund, Sweden}

\date{\today}

\begin{abstract}
In this article, we experimentally demonstrate a new way of controlling the group velocity of an optical pulse by using a combination of spectral hole burning, slow light effect and linear Stark effect in a rare-earth-ion-doped crystal. The group velocity can be changed continuously by a factor of 20 without significant pulse distortion or absorption of the pulse energy. With a similar technique, an optical pulse can also be compressed in time. Theoretical simulations were developed to simulate the group velocity control and the pulse compression processes. The group velocity as well as the pulse reshaping are solely controlled by external voltages which makes it promising in quantum information and quantum communication processes. It is also proposed that the group velocity can be changed even more in an Er doped crystal while at the same time having a transmission band matching the telecommunication wavelength. 

\end{abstract}

\maketitle

\section{INTRODUCTION}

Controlling the group velocity of light has attracted more and more attention due to its potential application in data synchronization, tunable optical buffer in optical communication \cite{Ramaswami2009}, optical information processing and optical switching \cite{Hamilton2002}. To date, varies techniques of controlling the group velocity of a pulse have been studied, such as electromagnetic induced transparency (EIT) \cite{Phillips2001, Bajcsy2003}, coherent population oscillation (CPO) \cite{bigelow2003}, stimulated Raman scattering (SRS) \cite{Sharping2005}, and stimulated Brillouin scattering (SBS) \cite{Shi2009}. More recently, group velocity control was demonstrated in silicon based resonators \cite{liu2008, xu2006}, with the target for on-chip application. However, the delay there is only typically in the range of hundreds of picoseconds. One major drawback of the above mentioned methods for group velocity control is their needs of a simultaneous optical pump beam which might introduce some background noise to the system, that could be especially detrimental in background sensitive single photon applications. In this article, we present a new technique where the group velocity of an optical pulse is solely controlled by an electric field. This is a desired feature when working with weak light situations, for example in quantum information and quantum communication processes. 

\begin{figure}[htbp]
\centering
\includegraphics[width=\linewidth]{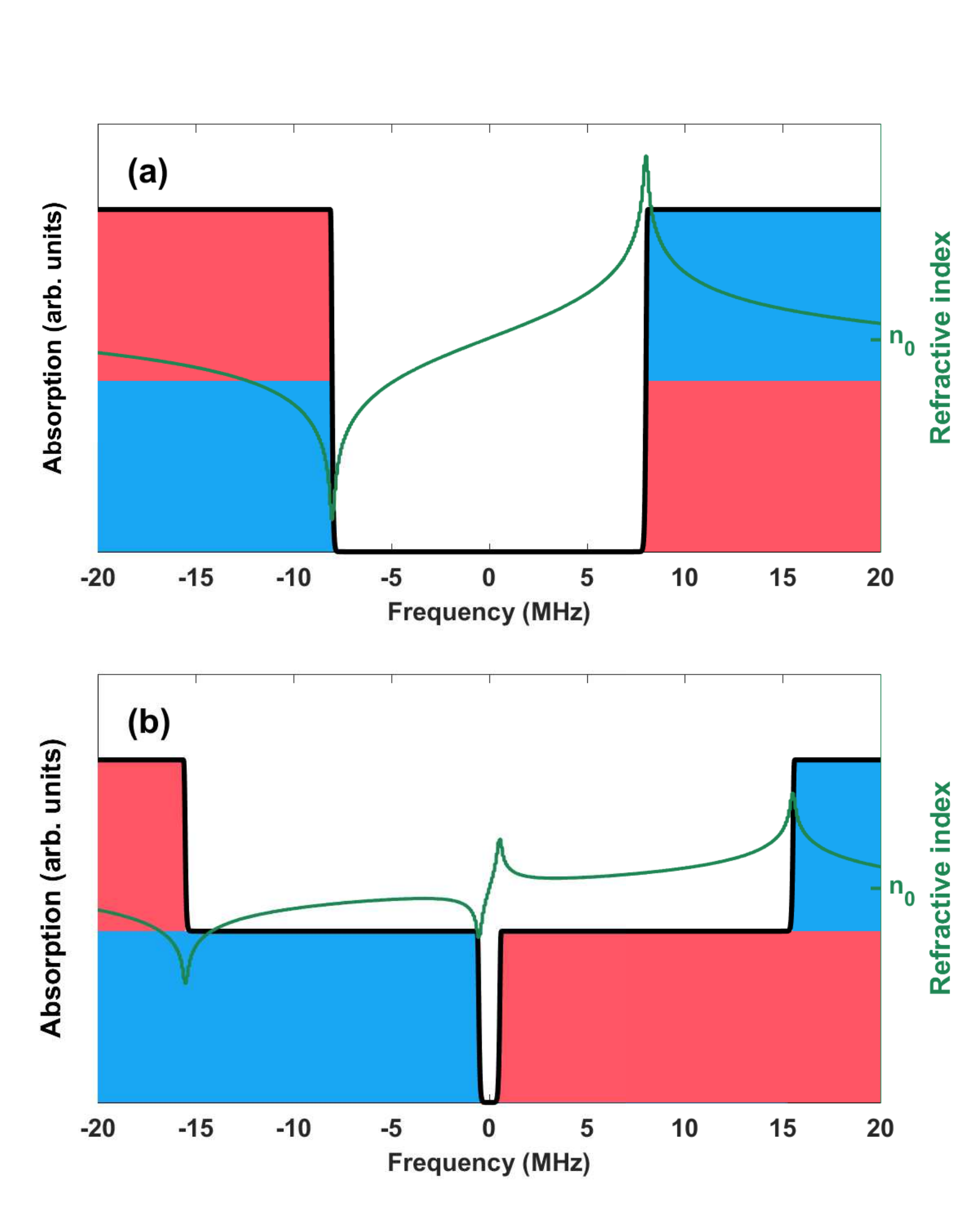}
\caption{A simplified sketch of the absorption structure, as well as the dispersion of the refractive index, where the red and blue color represents ions with opposite sign of their Stark coefficients while the black line shows the overall absorption. The green trace shows the real part of the refractive index over the frequency range, where n$_0$ stands for the refractive index of the host material and in this case n$_0$ = 1.8. (a) the original 16 MHz hole, no external voltages applied; (b) the 1 MHz spectral hole structure when an external voltage of 40 V is applied. The width of the hole as well as the slope of the dispersion changes as the external voltage is changed, which in return changes the group velocity.} 
\label{fig:filter_sketch}
\end{figure}

Our technique of controlling the group velocity of an optical pulse uses the combination of spectral hole burning induced slow light effect and Stark effect and can be described as follows.
A 16 MHz spectral hole was prepared in the center of the inhomogeneous broadening of a rare-earth-ion-doped crystal, with the target of making the edges of the hole as sharp as possible. A principle sketch of the final hole structure together with the real part of the refractive index is shown in Fig. \ref{fig:filter_sketch} (a). The blue and red colors represent ions with positive and negative Stark coefficients, respectively. If a voltage is applied across the crystal, the resonance frequencies of the 'blue' and 'red' ions will shift in different directions, which leads to a narrower hole. The change of the hole width is proportional to the applied external voltage since the frequency shift of the ions is proportional to the applied voltage according to the linear Stark shift. For an applied voltage of 40 V, a sketch of the new structure, as well as the real part of the refractive index are shown in Fig. \ref{fig:filter_sketch} (b), where in the vicinity of the left side of the hole there are only 'blue ions' while in the vicinity of the right side of the hole there are only 'red ions'. The dispersion across the transmission window is much steeper than that in Fig. \ref{fig:filter_sketch} (a). This causes the group velocity of an optical pulse to be much slower when propagating in the transmission window shown in Fig. \ref{fig:filter_sketch} (b) than when propagating in a transmission window as shown in Fig. \ref{fig:filter_sketch} (a) (See Eq. (\ref{eq:vg1}) and Eq. (\ref{eq:vg3}) in the Group velocity section for further information).

When a 10 $\mu$s long pulse with a Gaussian frequency profile is sent into the crystal, the group velocity of the pulse could be changed continuously by a factor of 20 by changing the external voltages applied across the crystal and the group velocity decreases monotonically as the external field is increased. Furthermore, an optical pulse can be reshaped by changing the hole width while the pulse is still inside the crystal. A 1 $\mu s$ long pulse was sent into the crystal at the presence of an external voltage of 40 V, which corresponds to a hole width of about 1 MHz. The group velocity of the pulse was small enough that almost the entire pulse was accommodated in the crystal. The electric field was then decreased rapidly, increasing the group velocity. In this way the first part of the pulse propagated a longer distance inside the crystal with a slower speed than the later part of the pulse, which caused the pulse transmitted from the crystal to be compressed in time.

Although the present experimental demonstration was carried out in Pr:Y$_2$SiO$_5$, it can be used in other rare-earth-ion-doped materials where spectral hole burning are possible and Stark or Zeeman effects can be applied to change the hole width. For example in Er doped crystals (e.g., Er:Y$_2$SiO$_5$). Er ions has an optical transition around 1.5 $\mu m$, at the telecommunication wavelength, where the fiber loss is minimum. The Stark coefficient is about 20 kHz/(V$\cdot$ cm$^{-1}$) in Er:YAlO3 \cite{Wang1992}, and 25 kHz/(V$\cdot$ cm$^{-1}$) in Er:LiNbO3 \cite{Hastings-Simon2006}. The ground state hyperfine splitting for Er ions is much wider than that of Pr ions \citep{Guillot2006,Baldit2010}, a spectral hole of 575 MHz can be prepared in Er:Y$_2$SiO$_5$ according to the hyperfine splittings provided in Ref. \cite{Milos_arXiv}. Since the tuning range of the group velocity is only limited by the widest hole width that can be created and the external voltages applied, the tuning range of the group velocity can be much larger in Er than that in Pr.

\section{GROUP VELOCITY} \label{vg}
The group velocity, $\mathbf{v}_g$, of an optical pulse can be calculated via the following equation,
\begin{equation}\label{eq:vg1}
\mathbf{v}_g = \frac{c}{n+\nu\frac{dn}{d\nu}}
\end{equation}
where c is the speed of light in vacuum, \textit{n} is the real part of the refractive index (for the phase velocity) and $\nu$ is the frequency of the light pulse. When an optical pulse propagates inside a dispersive medium, in our case inside a narrow transmission window, the real part of the refractive index can change dramatically, therefore, $\nu\frac{dn}{d\nu}$ can be much higher than \textit{n} itself, which in the present experiments can decrease the group velocity by four or five orders of magnitude \cite{Sabooni2013, Li2016}. 

Another way of interpreting the group velocity reduction is from the light-matter interaction perspective. 
As shown in Refs. \cite{Courtens1968, Shakhmuratov2010, Lauro2009} (note that in Ref. \citep{Shakhmuratov2010, Lauro2009}, the assumption is made that the material has a refractive index of n = 1), the group velocity can be expressed as a temporal energy storage process of the optical pulse energy in the absorbing ions via off-resonance interaction,
 \begin{subequations}
 \begin{align}
 \mathbf{v}_g = \frac{c/n}{1+\frac{U_{c}}{U_{em}}} \label{eq:vg2a}
 \end{align}
where $U_c$ is the energy density accumulated in the excitation of the off resonant centers and $U_{em}$ is the energy density of the electromagnetic field propagating with phase velocity c/n in the medium in the absence of dispersion. The equation shows that, the more energy is temporarily stored in the resonance centers, the lower the group velocity of the light. 
 
In fact, the refractive index, n, also originates from the off-resonant interaction but between light and the host material. Therefore, when light propagates inside a non-dispersive medium, the energy density $U_{em}$, also contains two parts, one is the energy density of the pure electromagnetic wave, $U_{vac}$ and another one is the energy density accumulated in the host atom, $U_{host}$. Similar as in Ref. \cite{Courtens1968}, the energy flow across a unit cross section inside the crystal per unit time is $ \mathbf{v}_g (U_{vac} + U_{host}+U_c)$, which should be the same as the energy flow through a unit cross section per unit time outside the crystal, $ c U_{vac}$. Then the group velocity can be expressed as
 \begin{align}
 \mathbf{v}_g = \frac{c}{1+\frac{U_{host}}{U_{vac}}+\frac{U_{c}}{U_{vac}}} \label{eq:vg2b}
 \end{align}
hence,
 \begin{equation}
\mathbf{v}_g = \frac{c}{1+\frac{U_{med}}{U_{vac}}} \label{eq:vg2c}
 \end{equation}
 \end{subequations}
where $U_{med} = U_{host} + U_{c}$ is the energy density accumulated in the medium which contains the effect from the host atoms and the resonant centers. In the case of no dispersion, the group velocity equals the phase velocity and we get $n = 1+\frac{U_{med}}{U_{vac}}$. Thus the higher the refractive index is, the stronger the interaction between the light and the material, hence the more energy is stored inside the material. To verify this, we can take a non-dispersive material with refractive index of \textit{n}, assume that we have a light beam with a certain amplitude and normal incidence enters a material with refractive index of n from vacuum, by comparing the energy density inside and outside the material (denoted $U_{inside}$ and $U_{vac}$, respectively) from the boundary condition, we get $U_{inside} = n U_{vac}$. If we make the same analogy as Courtens did in Ref. \cite{Courtens1968} that the energy density inside a material can be written as $U_{inside} = U_{med}+U_{vac}$, we get that $n = 1+\frac{U_{med}}{U_{vac}}$. We note that, Eq. (\ref{eq:vg2a}) is based on the assumption that the dispersion is due only to resonance centers other than the host material, while Eq. (\ref{eq:vg2c}) is suitable for any general case. That is with and without dispersion and where the dispersion can come from the host atoms or other resonance centers in the material or from both of them. Therefore, when an optical pulse enters a medium, a certain portion of the pulse energy is stored in the off-resonant ions (atoms) of the medium, and the Bloch vectors of these ions (atoms) are lifted up slightly from their ground state position.These ions will rephase at a later time, their Bloch vectors will point straight down again and the energy is returned back to the optical field. Intuitively, one can imagine that this process causes the light to propagate slower inside the medium.

In the case where the group velocity reduction is mainly caused by the spectral hole burning in an absorption profile such as our case here, the group velocity, $\mathbf{v}_g$, can be approximately calculated by \cite{Shakhmuratov2005, Walther2009, Sabooni2013},
\begin{equation}\label{eq:vg3}
\mathbf{v}_g \approx \frac{2\pi\Gamma}{\alpha}
\end{equation}
where $\Gamma$ is the width of the transmission window and $\alpha$ is the absorption coefficient outside the transmission window. 

Equations (1) - (3) can all be used to describe the group velocity but they emphasize different aspects. Eq. (\ref{eq:vg1}) is the formula to calculate the group velocity when one knows the dispersion, while Eq. (\ref{eq:vg2b}) and Eq. (\ref{eq:vg2c}) are ways to interpret the physical process of group velocity for the so called material slow light case \cite{Boyd2011, Courtens1968, Li2016}. Equation (\ref{eq:vg3}) provides an approximate but convenient estimation of the group velocity inside a spectral hole and is especially useful in our case here for making the first estimation of the tunability of the group velocity since it is proportional to the width of the transmission window and inversely proportional to the absorption coefficient. It is then straightforward to see that the tunablity of the group velocity is proportional to the tuning range of the transmission window. 

\section{EXPERIMENT}

The experiment was performed on a $6 \times 10 \times 10$ mm (crystal axes $b \times D1\times D2$ ), 0.05\% doped Pr:Y$_2$SiO$_5$ crystal at $\sim$ 2 K. The top and bottom surfaces of the crystal were coated with two sets of gold electrodes indicated as the red and black lines on the crystal in Fig. \ref{fig:setup} (a) (a photo of the crystal can be found in the Appendix of Ref. \cite{Li2016}). The energy levels involved in this experiment belong to the $^3H_4$ $ - $ $^1D_2$ transition centered around 605.978 nm with an inhomogeneous broadening of about 5 GHz and a homogeneous broadening of about 3 kHz. The hyperfine lifetime of the Pr$^{3+}$ ion ground state levels is about 100 s and can be extended to about 30 min at the presence of a weak (0.01 T) magnetic field\cite{Ohlsson2003}. 

A simplified setup is shown in Fig. \ref{fig:setup} (a), where a laser beam from a frequency stabilized Coherent 699 - 21 ring dye laser tuned to the center of the inhomogeneous broadening was split into two parts by a 90/10 beam splitter, where the stronger beam was focused onto the center of the crystal and used for the spectral structure preparation and characterization and then measured by a Thorlabs PDB150A photodetector (PD2), while the weaker one was directly measured by another Thorlabs PDB150A photodetector (PD1) and used as a reference beam to calibrate intensity variations. 

 \begin{figure}[htbp]
\centering
 \includegraphics[width=0.45\textwidth,natwidth=610,natheight=642]{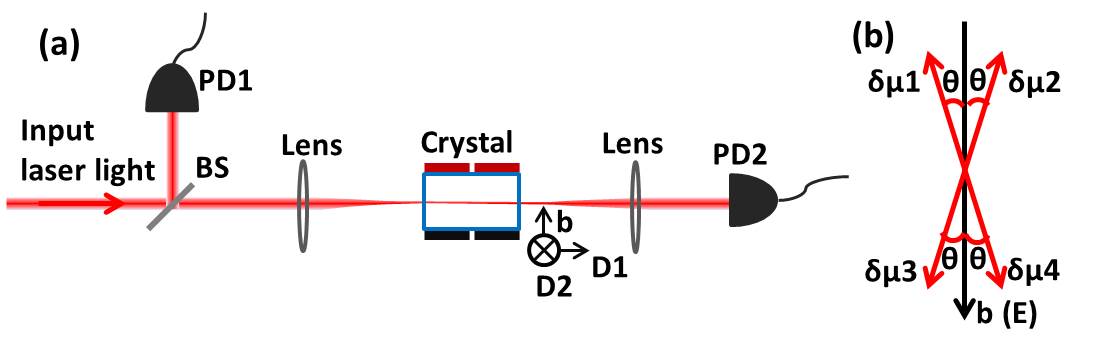}
\caption{Experimental setup. Laser light polarized along the $D2$ axis of the crystal is divided into two beams by a 90/10 beam splitter (BS), where the stronger beam is focused and directed along the center line of the crystal and recorded by a photodetector (PD2) while the weaker one is recorded by another photodectector (PD1) and used as a reference beam to calibrate any intensity fluctuations of the laser. (b) The permanent electric dipole moment difference between excited state and ground state of the Pr$^{3+}$ ($\boldsymbol{\delta\mu}1,2,3,4$) relative to the electric field applied (\textbf{E}). The electric field is along the b axis of the crystal and the magnitude of the projection of the dipole moment difference onto the electric field direction is the same for all the 4 possible $\boldsymbol{\delta\mu}$ ($\theta = 12.4^{\circ}$).} 
\label{fig:setup}
\end{figure}

The permanent electric dipole moment of the ground state is different from that of the excited state for Pr$^{3+}$ ions and there are four possible orientations for the dipole moment difference \cite{graf1997, Li2016}. When an external voltage is applied across the crystal, the resonance frequency of the ions will be Stark shifted by $\Delta_s$, where 
\begin{equation}\label{eq:StarkShift}
\Delta_s = \frac{\boldsymbol{\delta\mu} \cdot \mathbf{E}}{\hbar} = \pm \frac{\lvert \boldsymbol{\delta\mu} \rvert \mbox{cos}(\theta)}{\hbar} \cdot \frac{\Delta V}{d} 
\end{equation}
where $\hbar$ is the reduced Planck constant, $\boldsymbol{\delta\mu}$ is the difference between the excited and ground state dipole moments for $Pr^{3+}$ ions, \textbf{E} is the electric field across the crystal (along the b axis), $\theta$ is the angle between the electric field and $\boldsymbol{\delta\mu}$, $\Delta V$ is the difference in the electric potential of the top and bottom electrodes, and \textit{d} is the distance between these two electrodes which in our experiment is 6 mm. The four possible $\boldsymbol{\delta\mu}$ of Pr$^{3+}$ are oriented at an angle of $\theta = 12.4^{\circ}$ relative to the b axis of the crystal as shown in Fig. \ref{fig:setup} (b). The magnitude of $\boldsymbol{\delta\mu}/\hbar$ is about 111.6 kHz/(V cm$^{-1}$) \cite{graf1997}. Since the \textbf{E} field is applied along the b axis and because of the symmetry of $\boldsymbol{\delta\mu}$, there will be two effective Stark coefficients for Pr$^{3+}$ with the same magnitude but opposite signs. At the presence of a certain electric field, the resonance frequency of half of the ions, those with positive Stark coefficient (referred as 'blue ions'), will shift to higher frequencies while the rest of the ions, those with negative Stark coefficient (referred to as 'red ions'), will shift to lower frequencies.


Originally, a 16 MHz square transmission window was created using optical pumping. However, due to the inhomogeneity of the \textbf{E} field inside the crystal (see the Appendix in Ref. \cite{Li2016}) and the possible small mismatch of the electric field with the b axis of the crystal, when an electric field is applied to change the hole width, different ions sees slightly different electric field and shift slightly different amount in frequency, which smears out the sharpness of the hole edge and induces some absorption close to the edge of the hole. To compensate for this effect, the actual hole burning sequence was optimized in the following way, 16 MHz, 8 MHz, 1 MHz and 500 kHz hole burning sequences were employed consecutively at the presence of a 0 V, 20 V, 40 V, and 41.5 V external voltages, followed by the same 16 MHz hole burning at 0 V in the end. This process is repeated 50 times until a hole with high transmittance and sharp edges was achieved. The sharpness of the hole was optimized by sending a frequency chirped readout pulse after the hole creation and examine the coherent beating signal at the first hole edge. The higher the beating amplitude, the sharper the hole edge. The Rabi frequency of the hole burning sequence was kept low to avoid any power broadening and instantaneous spectral diffusion \cite{Liu1990,Equall1995} in order to make the hole edge as sharp as possible. The whole process took about half a minute, so a 0.01 T magnetic field was applied to decrease the hyperfine relaxation rate such that the hole did not degrade much during the preparation process. A simplified sketch of the final hole structure as well as the structure at the presence of an external voltage of 40 V together with the real part of the refractive index can be found in Fig. \ref{fig:filter_sketch}.


\section{RESULT AND DISCUSSION}


\subsection{Group velocity control}
\begin{figure}[htbp]
\centering
\includegraphics[width=\linewidth]{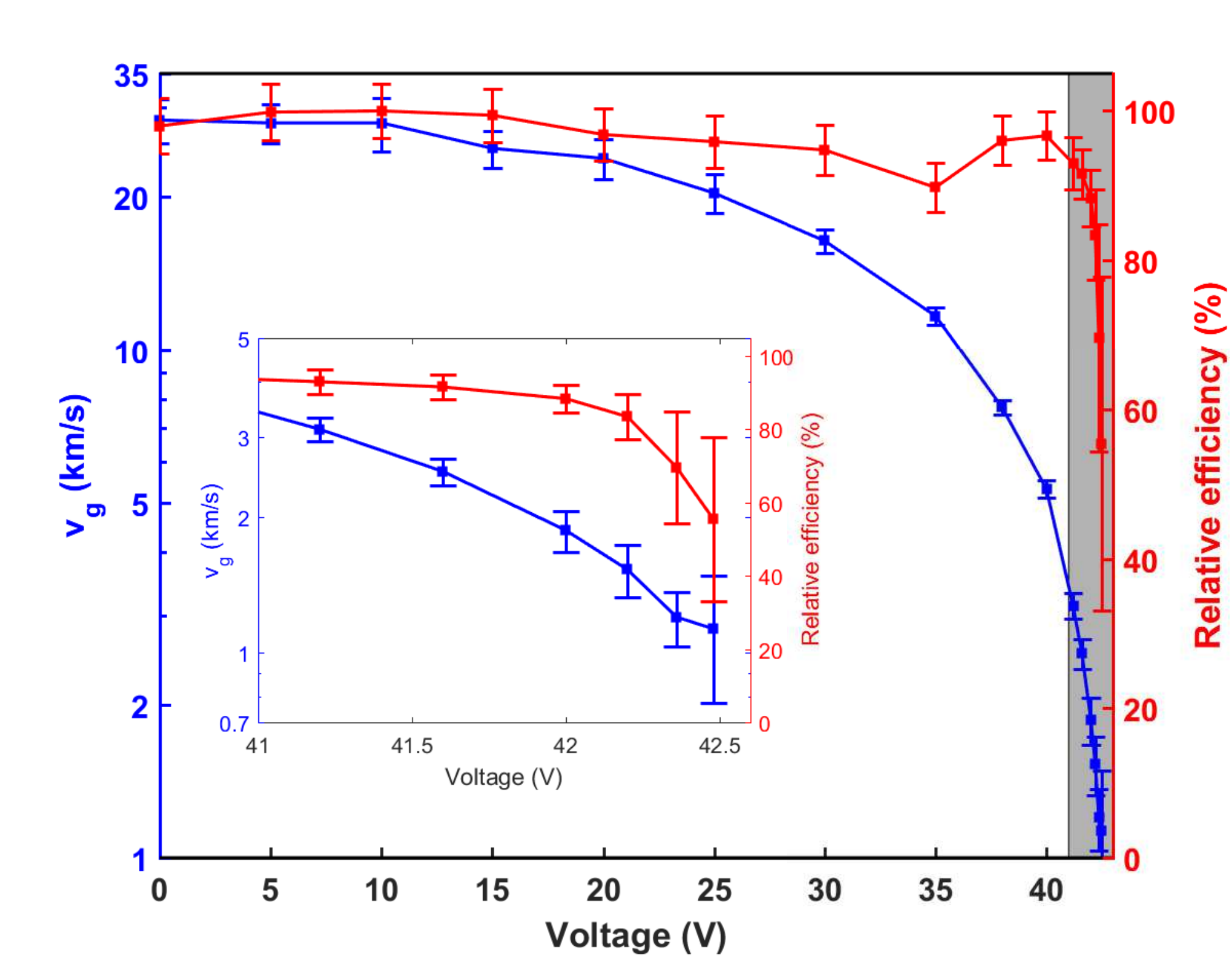}
\caption{The group velocity (blue) and the relative transmission efficiency (red) at different external voltages. The group velocity can be changed by a factor of 20 while keeping the relative efficiency more than 80\%.} 

\label{fig:vg_effi}
\end{figure}

After the structure preparation, 16 Gaussian pulses, each with a full width at half maximum (FWHM) of 10 $\mu s$ and a cutoff duration of 100 $\mu s$, were sent into the crystal successively while the external voltage applied across the crystal was increased monotonously for each probe pulse from 0 V to 42.5 V. A part of the incoming pulse (reference beam) and the transmitted pulse were measured by PD1 and PD2 shown in Fig. \ref{fig:setup} (a). By comparing the time delay of the transmitted signal and the reference beam, the group velocity of the optical pulses can be calculated for each electric field. The relative efficiency was calculated by comparing the area of the transmitted pulse at different electric fields with the area of the transmitted pulse with no external voltage (widest hole) after correction using the reference beam signal. The measurement was repeated 150 times and the result is shown in Fig. \ref{fig:vg_effi} with the dot represents the average value for each voltages and the bars associated with them are one standard deviation away from the average. The gray area was zoomed in as an inset. The group velocity of the probe pulse changed from ~30 km/s to ~1.5 km/s (when external voltages were changed from 0 V to 42.2 V) while still having more than 80\% relative transmittance. Beyond this, the absorption increases dramatically as the hole width becomes comparable to the frequency width of the pulse. 1 km/s was achieved with about 50\% transmittance.

We can interpret the reduction of the group velocity from the three aspects that were discussed in section \ref{vg}. From Eq. (\ref{eq:vg1}), it is clear that the steeper the dispersion, the slower the group velocity. When an external voltage is applied across the crystal the hole becomes narrower, and the slope of the dispersion becomes steeper as shown in Fig. \ref{fig:filter_sketch}. Hence $\nu\frac{dn}{d\nu}$ becomes much bigger, which decreases $\mathbf{v}_g$ greatly.

As revealed from Eq. (\ref{eq:vg2c}), the more energy stored in the medium, the slower the group velocity of the light. When the hole becomes narrower, the resonant frequency of the absorbing ions are much closer to the probe frequency, therefore, the off-resonance interaction between the light and the ions becomes much stronger, which leads to more energy stored in the off-resonant ions and the group velocity decreases. For the 16 MHz hole, with group velocity of $\mathbf{v}_g = 30$ km/s, according to Eq. (\ref{eq:vg2c}), the fraction of energy temporally stored in the crystal is $99.98\%$, while for a group velocity of $\mathbf{v}_g = 1.5$ km/s, the fraction of energy stored in the crystal is $99.9991\%$. 

We can easily estimate the tunability of the group velocity of the structure from Eq. (\ref{eq:vg3}), when the width of the hole changes from 16 MHz (0 V) to 0.25 MHz (estimated experimental hole width at the bottom of the hole at 42.2 V), a reduction of a factor of $\frac{16}{0.25}*\frac{1}{2} = 32$ should be achieved (the $\frac{1}{2}$ comes from the fact that the absorption coefficient just outside the hole changes by a factor of 2 for these two structures, see Fig. \ref{fig:filter_sketch}). However, experimentally we only get a factor of 20 in group velocity reduction, which is probably due to the decrease in the sharpness of the hole edge, so even though the window of full transmission is only 0.25 MHz, the full width at half maximum of the hole is actually much wider than that. 

Concerning the absolute efficiency, it has been shown previously that the remaining absorption inside a 18 MHz spectral hole of a Pr:Y$_2$SiO$_5$ crystal with the same doping concentration as in the present experiment can be less than 0.1 dB/cm \cite{Sabooni2013}. In our case this will correspond to an absolute transmission efficiency of higher than 97\% for the 16 MHz spectral hole. We can then renormalize the relative efficiency according to this number to estimate the absolute efficiency for each group velocity. 

In principle, as shown in Eq. \ref{eq:vg3}, the tunability of the group velocity is only limited by the tunable range of the spectral hole width. Therefore, it could be greatly expanded if a much wider hole can be prepared from the beginning. For example, in Er:Y$_2$SiO$_5$, a spectral hole that is 575 MHz can be prepared given the hyperfine structures of the ground and excited states \cite{Milos_arXiv}. Therefore instead of a factor of 20 demonstrated here, a factor of 600  could be achieved ideally. Besides this, the absolute group velocity can be changed by changing the absorption coefficient at the hole edges and this can be easily changed by changing the laser frequency to the side of the inhomogeneous  absorption profile for the structure preparation process (or changing to another crystal with higher or lower absorption coefficient). The closer to the center of the absorption, the lower group velocity one can get while keeping the same tunability.

 The bandwidth-delay product, $\mathbf{BD}$, for such a slow light structure can be approximately calculated by \cite{Lauro2009}, 
\begin{equation}\label{eq:TB}
\mathbf{BD} = \frac{L}{\mathbf{v}_g}\Gamma \approx \frac{\alpha L}{2\pi}
\end{equation}
where \textit{L} is the length of the crystal. Since the absorption coefficient can easily be changed by going to another wavelength and the length of the crystal can be altered according to the purpose of the application, the bandwidth-delay product can be therefore easily changed. A more detailed discussion can be found in Refs. \cite{Boyd2005, Sabooni2013}.

\subsection{Pulse compression}
\begin{figure}[htbp]
\centering
\includegraphics[width=\linewidth]{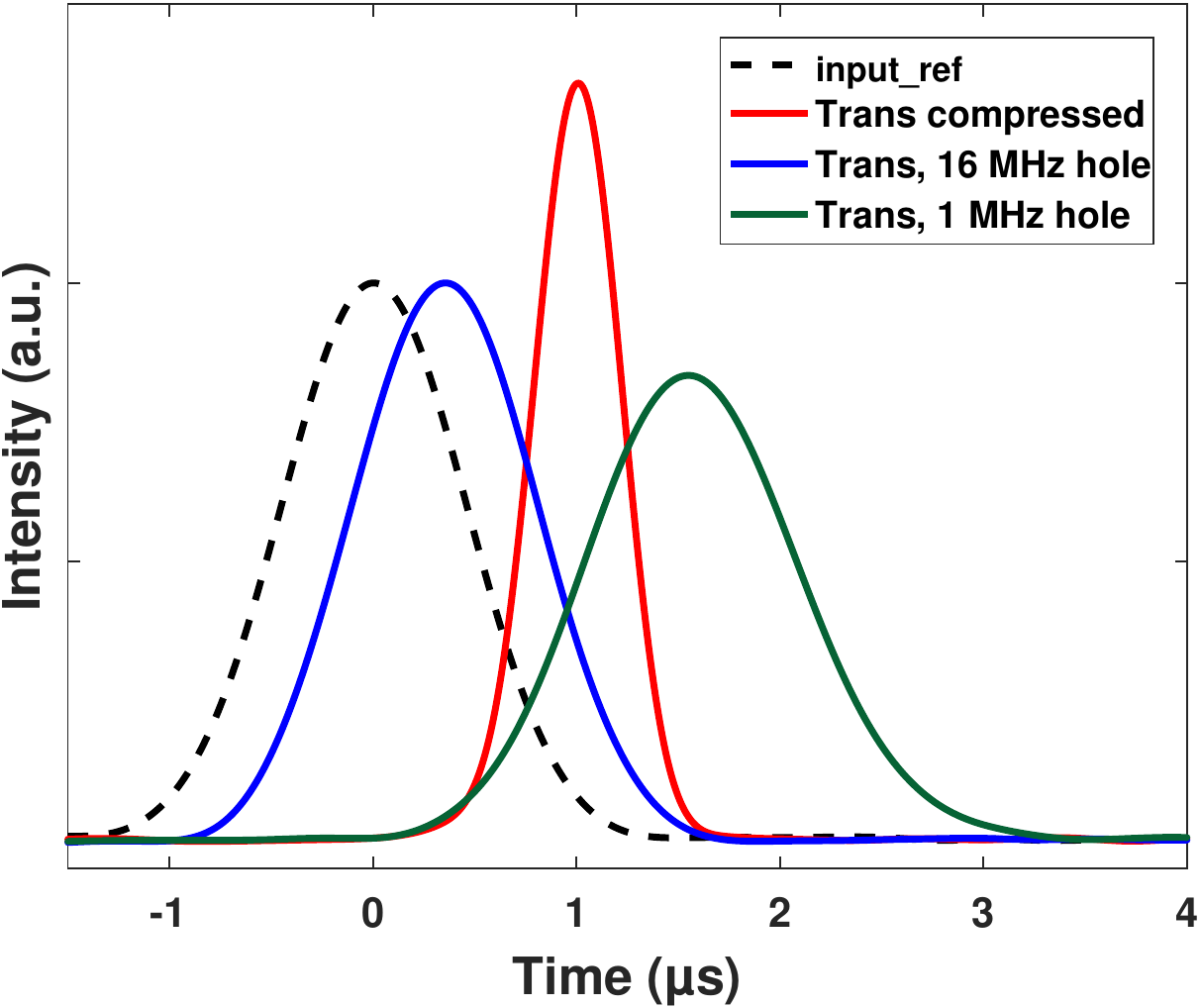}
\caption{Pulse compression. Dashed black trace: the reference incoming pulse; the blue trace: 0 V; the dark green one: 40 V; the red trace: 40 V changed to 0 V just before the pulse is about to exit the crystal. } 

\label{fig:pulsecompression}
\end{figure}

Group velocity control can also be used to reshape an optical pulse in time. After the creation of the 16 MHz hole (structure shown in Fig. \ref{fig:filter_sketch} (a)), a 1 $\mu s$ long pulse was sent into the crystal at time t = 0, shown as the black dashed trace in Fig. \ref{fig:pulsecompression}. When no voltage is applied, the pulse comes out of the crystal at time $\tau_1$ = 350 ns, shown as the blue trace, and the group velocity of the pulse is $\mathbf{v}_{g1} = L/\tau_1 \approx 29$ km/s, while when the pulse enters the crystal at the presence of an external voltage of 40 V (structure shown as in Fig. \ref{fig:filter_sketch} (b)), it propagates with a much lower group velocity, $\mathbf{v}_{g2} \approx 6.5$ km/s, and comes out of the crystal at $\tau_2 = 1.54$ $\mu s$, shown as the green trace in Fig. \ref{fig:pulsecompression}.

Because of the low group velocity when 40 V was applied across the crystal, the 1 $\mu s$ long light pulse  was greatly compressed specially, from 300 m long in vacuum to about 6 mm long inside the crystal. Therefore, almost the entire pulse can be accommodated inside the crystal. If the external voltage is switched off when the pulse is just about to exit, then the first part of the pulse will go through the entire crystal with a group velocity of $\mathbf{v}_{g2} = L/\tau_2$, while the last part of the pulse will initially propagate with group velocity $\mathbf{v}_{g2}$ and change to a group velocity of $\mathbf{v}_{g1}$ after the E field is switched off. Therefore, the last part of the pulse spends much shorter time inside the crystal and the pulse will be compressed in time, shown as the red trace in Fig. \ref{fig:pulsecompression}. 

To estimate how much a pulse can be compressed in time, assume that the pulse propagates with a group velocity of $\mathbf{v}_{g2}$ at first and the pulse length is $l = \mathbf{v}_{g2} \tau$, and the crystal is long enough to accommodate the whole pulse length. If the group velocity is changed to $\mathbf{v}_{g1}$ instantaneously just when the first part of the pulse is about to exit, it takes $\tau^{\prime}=l/\mathbf{v}_{g1}$ for the last part of the pulse to exit the crystal. Therefore, a pulse initially with a duration of $\tau$ will be changed to $\tau^{\prime} = \frac{\mathbf{v}_{g2}}{\mathbf{v}_{g1}}\tau$. Therefore, the compression factor, defined to be the ratio of the width of the original pulse to the width of the compressed pulse in time, would be $\frac{\tau}{\tau^{\prime}} = \frac{\mathbf{v}_{g1}}{\mathbf{v}_{g2}}$. Hence, the bigger the ratio of the two group velocities, the more the pulse can be compressed. In reality, the electric field can not be changed instantaneously, so the compression factor will be smaller.  However, the compression factor will always be $\frac{\mathbf{v}_{g2}}{\mathbf{v}_{g1}}$, if the whole pulse is inside the crystal before and after changing the E field. 

If we look at the pulse compression process from the energy density point of view, the pulse first propagate in a structure as in Fig. \ref{fig:filter_sketch} (b), where the group velocity is low, the energy of the optical pulse is mainly stored in the close-by off-resonant ions. When the electric field is decreased, those ions will shift outwards relative to the hole center, constitute a wider hole, and emit light at another frequency, so the pulse coming out would have a wider frequency distribution compared to the pulse sent in. The wider frequency span may be viewed as enabling the pulse compression in time.

A relevant quantity about the pulse compression is the efficiency of the compression process, which can be calculated by comparing the pulse area of the compressed pulse with the uncompressed pulse from Fig. \ref{fig:pulsecompression}. If the transmitted energy for a 16 MHz hole (blue trace)is set to 100\%, then the transmitted energy for the 1 MHz hole is about 96.6\% while the energy transmitted for the compressed pulse is about 66\%. To evaluate the energy loss during the pulse compression process, a Max Bloch simulation was performed and the detailed information about the simulation can be found in the Simulation section of Ref. \cite{Li2016}. Two different simulations were carried out for comparison. Case I corresponds to the actual experiment when a 1 $\mu$s pulse is sent into the 1 MHz transmission window as shown in Fig. \ref{fig:filter_sketch} (b). Before the pulse exits, the E field is switched off and the transmission window changes to 16 MHz, just as shown in Fig. \ref{fig:filter_sketch} (a). In this case, if we look at the right hand side of the hole structure there only exist 'red' ions before the field changes so the energy will mainly stored in the 'red' ions, while the 'red' and 'blue' ions overlaps after switching off the E field. Therefore, when light was re-emitted by the 'red' ions, it might be resonantly absorbed by the 'blue' ion. We expect this will cause some energy loss. Therefore, case II is simulating a similar situation but on the right side of the hole all the ions are 'red' ions while on the left side of the hole all the ions are 'blue' ions as the structure shown in Fig. \ref{fig:filter_sketch} (b) if the top half part is neglected. Thanks to this arrangement in the simulation, there will never be any overlap of the resonance frequency of the two groups of ions. Hence by comparing these two cases, the energy loss due to the overlap in resonance frequency of the two groups of ions can be mapped out. 

\begin{figure}[htbp]
\centering
\includegraphics[width=\linewidth]{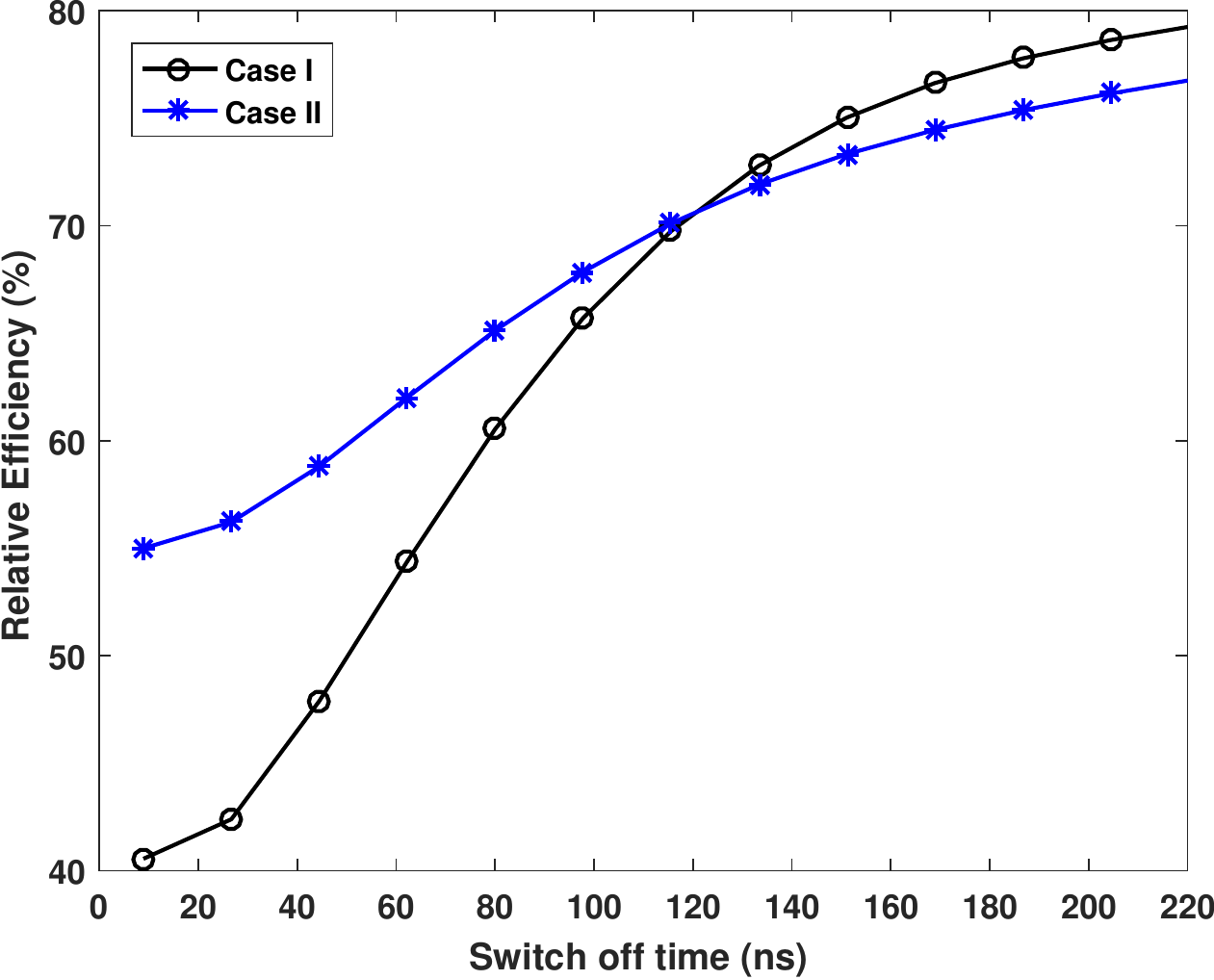}
\caption{Pulse compression efficiency as a function of E field switch off time. The blue circle are from simulation case I while the red stars are from simulation case II. See main text for the description of the two cases.} 
\label{fig:simulation1}
\end{figure}

\begin{figure}[htbp]
\centering
\includegraphics[width=\linewidth]{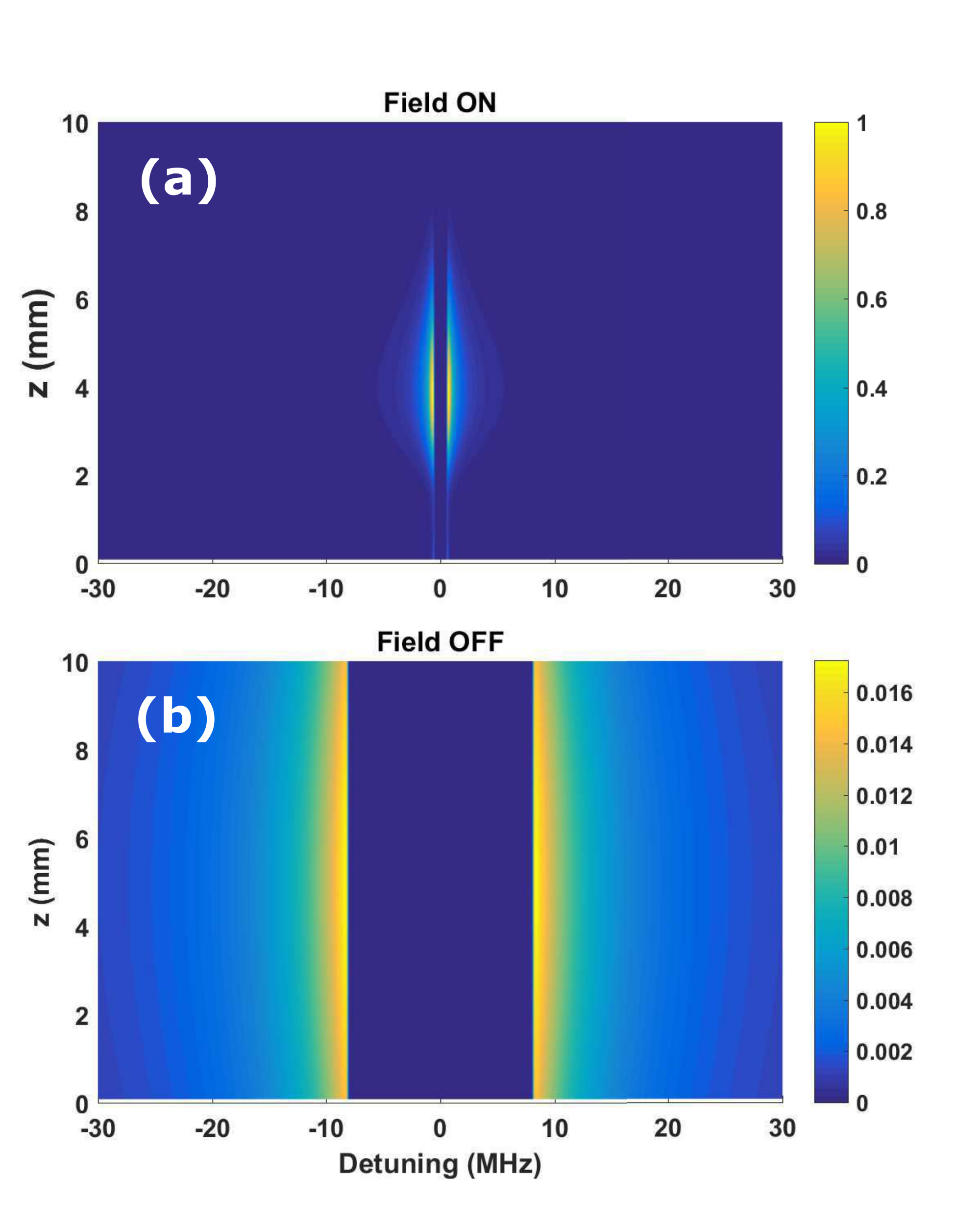}
\caption{Energy storage for different hole structures. (a) The energy distribution when the hole is 1 MHz and (b) The energy distribution when the hole is 16 MHz. 1 represents maximum stored energy in the ions while 0 represents no energy stored in the ions. Note that the scale of the color bars are different in (a) and (b). As can be seen, the energy is much more confined both in frequency and space when the pulse propagates slowly in the 1 MHz hole. } 

\label{fig:simulation2}
\end{figure}

The simulated efficiency of the pulse compression process as a function of the switch off time of the E field can be found in Fig. \ref{fig:simulation1}. It shows that the slower the E field is switched off, the higher the efficiency and when the E field was switched off almost instantaneously, the energy loss are 40.5\% for case I and 55\% for case II, respectively. The 15 \% lower efficiency in case I can be attributed to the overlap of the two groups of ions ('red' and 'blue') as discussed above. 
However, even for case II where the ions never overlap with each other, the efficiency is still only 55 \% when the E field is switched off rapidly and this is probably due to the fact that the energy distribution of the pulse propagating inside a 1 MHz hole and a 16 MHz hole are quite different as shown in Fig. \ref{fig:simulation2}. When the E field is switched off almost instantaneously, the ions will Stark shift to another resonance frequencies accordingly and the hole changes from 1 MHz to 16 MHz. However, the resulting energy distribution doesn't match the 16 MHz structure, therefore, the rephasing process will not be complete and it is reasonable to assume that some energy would be lost due to this. As can be seen from Fig. \ref{fig:simulation1} that when the E field is switched off slower than 120 ns the energy loss due to this effect will decrease dramatically for both cases. 

In case I, if the E field is switched off in 200 ns, the relative efficiency one can reach is around 79 \% while in our experiment, only 66 \% was achieved. This could be due to the remaining absorption in the hole. As can be seen from Fig. \ref{fig:vg_effi}, there are certain absorption for each E field when the pulse propagates inside the hole, especially for an applied voltage of 35 V where the transmission efficiency is about 90 \%, which leads to some loss to the pulse energy. This could be minimized by further optimize the structure preparation process.

In principle this technique can be used to stretch the pulse in time as well by having a higher group velocity at the beginning and then change to a lower one. However, this would require a crystal that is long enough to fit the entire pulse when it propagates at a higher group velocity, therefore it is practically more difficult to implement. Nevertheless, one might be able to do this with a multipass beam arrangement in a crystal or in a slow light cavity\cite{Sabooni2013}. 

\section{CONCLUSION}

In conclusion, we demonstrate a device in which the group velocity of an optical pulse can be changed continuously by a factor of 20 solely by applying an external electric field. The energy loss of the transmitted pulse is kept below 20\% over the whole group velocity tuning range. The device can also be used to reshape and compress the pulse. The maximum compression of a pulse depends on the ratio of the maximum and minimum group velocity with which the pulse propagates inside the crystal. The device is based on creating a semi permanent transmission window in the absorption line of a rare-earth-doped crystal using optical pumping. By Stark shifting the resonance frequency of the ions, the structure of the transmission window as well as the dispersion of the refractive index across the transmission window will be changed, hence the group velocity of the pulse propagating inside the transmission window will be changed accordingly. It is proposed that the tunable range of the group velocity can be further extended in other rare-earth-doped crystals, such as Er$^{3+}$ doped crystals. As the group velocity is controlled purely by an external electric field, and does not require any additional optical pulses, it can be especially useful in weak light situations. For example, in quantum information and quantum communication processes. Moreover, since it is a solid state device it can be integrated with other devices and could open up on-chip applications.
 
\section*{ACKNOWLEDGMENTS}
This work was supported by the Swedish Research Council, the Knut $\&$ Alice Wallenberg Foundation. The research leading to these results also received funding from the People Programme (Marie Curie Actions) of the European Union's Seventh Framework Programme FP7 (2007-2013) under REA grant agreement no. 287252 (CIPRIS) and Lund Laser Centre (LLC).

\section*{APPENDIX}

As can be seen from the inset of Fig. \ref{fig:vg_effi} the error bars for the relative efficiency become much bigger when the applied voltages are higher than 42 V (which corresponds to a hole width of less than 300 kHz). This is attributed to the 50 Hz noise in the high voltage amplifier used for this experiment. As discussed before, the Stark shift of the ions is linearly proportional to the applied external field. The variation in the voltages applied to the crystal will cause variation in the Stark shift of the ions, which will affect the preparation process of the hole structure as well as the hole width when probing. When the width of the hole is comparable to the frequency spread of the optical pulse, a small fluctuation in the hole width can cause a dramatic change in the transmission as it changes exponentially with the number of ions inside the frequency width of the pulse. Therefore, the relative error for the transmission is much higher when the hole width is really narrow. 
\begin{figure}[htbp]
\centering
\includegraphics[width=\linewidth]{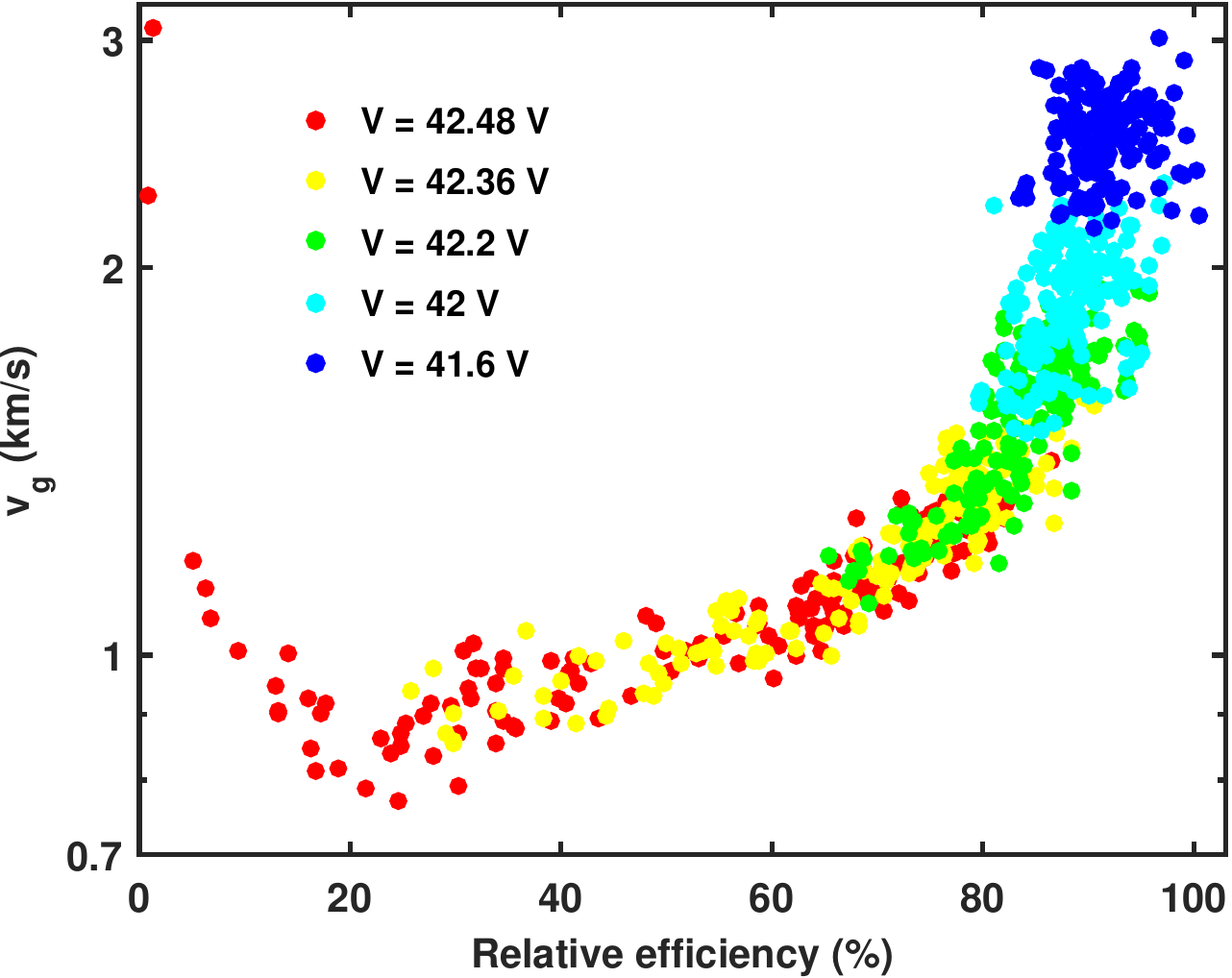}
\caption{The correlation of the group velocity and the relative efficiency at different external voltages.} 

\label{fig:Correlation}
\end{figure}

Since the fluctuation in the efficiency is caused by the fluctuation in the hole width, i.e., a lower efficiency is due to a narrower hole width. According to Eq. \ref{eq:vg3}, a narrower hole width gives a lower group velocity as well. Therefore, the group velocity and the relative efficiency should be correlated to each other. This can be verified from Fig. \ref{fig:Correlation} which shows the group velocity and relative efficiency of all the 150 measurement when the applied voltages are higher than 41.6 V (hole width narrower than 500 kHz). It shows that when the relative efficiency is higher than 20 \%, the group velocity is positively correlated with the relative efficiency, so that the lower the relative efficiency the lower the group velocity. However, when the relative efficiency drops below 20 \%, the group velocity and the relative efficiency shows an anti-correlation.

\bibliography{GroupVelocityControl}

\end{document}